\documentclass[10pt,letterpaper]{article}
\usepackage[colorlinks=true, hyperindex, breaklinks]{hyperref}
\usepackage[top=0.85in,left=1.75in,footskip=0.75in]{geometry}
 
\usepackage{amsmath,amssymb}

\usepackage{changepage}

\usepackage[utf8x]{inputenc}

\usepackage{textcomp,marvosym}

\usepackage{cite}

\usepackage{nameref,hyperref}

\usepackage{microtype}
\DisableLigatures[f]{encoding = *, family = * }

\usepackage[table]{xcolor}

\usepackage{array}

\newcolumntype{+}{!{\vrule width 2pt}}

\newlength\savedwidth

\raggedright
\usepackage[aboveskip=1pt,labelfont=bf,labelsep=period,justification=raggedright,singlelinecheck=off]{caption}

\makeatletter
\renewcommand{\@biblabel}[1]{\quad#1.}
\makeatother

\usepackage{lastpage,fancyhdr,graphicx}
\usepackage{epstopdf}
\fancyheadoffset[L]{2.25in}
\usepackage{verbatim}
\usepackage{fancyvrb}
\usepackage{xcolor}
\usepackage{dirtree}

\usepackage{listings}

\definecolor{darkpink}{rgb}{0.75,0.25,0.5}
\definecolor{darkgreen}{rgb}{0.18,0.54,0.34}
\definecolor{bg}{rgb}{0.95,0.95,0.95}

\usepackage{ulem}

\begin{document}
\vspace*{0.2in}

\begin{flushleft}
{\Large
\textbf\newline{Quantum++: A modern C++ quantum computing library} 
}
\newline
\\
Vlad Gheorghiu\textsuperscript{1,2,3*}
\\
\bigskip
\textbf{1} softwareQ Inc., Kitchener ON, Canada
\\
\textbf{2} Institute for Quantum Computing, University of Waterloo, Waterloo ON, Canada
\\
\textbf{3} Department of Combinatorics and Optimization, University of Waterloo, Waterloo ON, Canada
\\
\bigskip

%
%


* vlad@softwareq.ca

\end{flushleft}
\section*{Abstract}
Quantum++ is a modern general-purpose multi-threaded quantum computing library written in C++11 and composed solely of header files. The library is not restricted to qubit systems or specific quantum information processing tasks, being capable of simulating arbitrary quantum processes. The main design factors taken in consideration were the ease of use, portability, and performance. The library's simulation
capabilities are only restricted by the amount of available physical memory. On a
typical machine (Intel i5 8Gb RAM) Quantum++ can successfully simulate the
evolution of 25 qubits in a pure state or of 12 qubits in a mixed state
reasonably fast. The library also includes support for classical reversible logic, being able to simulate
classical reversible operations on billions of bits. This latter feature may be useful in testing quantum circuits
composed solely of Toffoli gates, such as certain arithmetic circuits.



\section{Introduction\label{sct::intro}}
Quantum computing is a disruptive technology that promises great benefits for a plethora of applications, ranging
from medicine and chemistry to machine learning and simulation of physical systems. However, today's most advanced quantum computers are not yet large enough for performing universal quantum computation, hence their applicability is 
still limited. Being able to simulate small sized quantum computers is therefore of paramount importance, as it allows 
the scientist or engineer to understand better the results she or he would expect from a quantum machine of similar size, as well as providing a better understanding of quantum computing itself. Below we describe a quantum computing library that can be used in research or exploratory work in quantum information and computation.

\href{https://github.com/vsoftco/qpp}{Quantum++}, available online at \url{https://github.com/vsoftco/qpp}, is a C++11 general purpose quantum computing library, composed solely of header files. It uses the \href{http://eigen.tuxfamily.org/}{Eigen~3} linear algebra library and, if available, the \href{http://openmp.org/}{OpenMP} multi-processing library. For additional \href{http://eigen.tuxfamily.org/}{Eigen~3} documentation see \url{http://eigen.tuxfamily.org/dox/}. For a simple \href{http://eigen.tuxfamily.org/}{Eigen~3} quick ASCII reference see \url{http://eigen.tuxfamily.org/dox/AsciiQuickReference.txt}. 

The simulator defines a large collection of (template) quantum computing related functions and a few useful classes. The main data types are complex vectors and complex matrices, which we will describe below. Most functions operate on such vectors/matrices passed by value and \emph{always} return the result by value, without ever mutating their arguments. The design is inspired from functional programming, where functions do not mutate their arguments and do not have side effects. Those design choices make the library ideal to use or integrate in multi-processing frameworks. Collection of objects are implemented via the standard library container \verb!std::vector<>!, instantiated accordingly. 

We decided to avoid using a complicated class hierarchy and focus on a functional style-like approach, as we believe the latter is more suitable for a relatively small API and allows the user to focus on the quantum algorithm design rather than on object-oriented design. In addition, there is absolutely no need for explicit memory allocations or usage of (raw) pointers. All allocations, initializations and release of resources are performed by the library, hence the user is not at risk of forgetting to de-allocate memory, use un-initialized objects, or overflowing buffers, which are the most common, dangerous and hard to diagnose mistakes in the world of C and C++ programming.

Although there are many available quantum computing libraries/simulators written in various programming languages, see~\cite{QCsims} for a comprehensive list, what makes \href{https://github.com/vsoftco/qpp}{Quantum++} different is the ease of use, portability and high performance. The library is not restricted to specific quantum information tasks, but it is intended to be multi-purpose and capable of simulating arbitrary quantum processes. We have chosen the C++ programming language (standard C++11) in implementing the library as it is by now a mature standard, fully (or almost fully) implemented by the most common compilers, and highly portable.

Other unique features of \href{https://github.com/vsoftco/qpp}{Quantum++} include the ability of simulating classical reversible networks up to billions of
bits (this feature may be useful in testing quantum circuits
composed solely of Toffoli gates, such as certain arithmetic circuits in e.g.~\cite{1706.06752}), strong multi-threading abilities, as well as built-in support for higher dimensional systems (qudits) that allows treating qubits and qudits on the same footing. 

In the reminder of this manuscript we describe the main features of the library, ``in a nutshell" fashion, via a series of simple examples. We assume that the reader is familiar with the basic concepts of quantum mechanics/quantum information. For a comprehensive introduction to the latter see e.g. \cite{NielsenChuang:QuantumComputation}. This document is not intended to be a comprehensive documentation, but only a brief introduction to the library and its main features. For a detailed reference see the manual available as a \verb!.pdf! file in \verb!./doc/refman.pdf!. For detailed installation instructions as well as for additional information regarding the library see the Wiki page at~\url{https://github.com/vsoftco/qpp/wiki}. If you are interesting in contributing, or for any comments or suggestions, please contact me.

\href{https://github.com/vsoftco/qpp}{Quantum++} is free software: you can redistribute it and/or modify it under the terms of the MIT
License \url{https://opensource.org/licenses/MIT}.

\section{Installation\label{sct::install}}
To get started with \href{https://github.com/vsoftco/qpp}{Quantum++}, first install the \href{http://eigen.tuxfamily.org/}{Eigen~3} library 
from \url{http://eigen.tuxfamily.org} into your home directory, as \verb!$HOME/eigen!. Here we implicitly assume that you use a UNIX-like system, although everything should translate into Windows as well, with slight modifications.
You can change the name of the directory, but in the current document we will use \verb!$HOME/eigen!
as the location of the \href{http://eigen.tuxfamily.org/}{Eigen~3} library. Next, download the \href{https://github.com/vsoftco/qpp}{Quantum++} library from~\url{https://github.com/vsoftco/qpp/} and unzip it 
into the home directory as \verb!$HOME/qpp!. Finally, make sure that your compiler supports
C++11 and preferably \href{http://openmp.org/}{OpenMP}. For a compiler we recommend \href{https://gcc.gnu.org/}{g++} version 5.0 or later or \href{http://clang.llvm.org}{clang} version 3.7 or later (previous versions of \href{http://clang.llvm.org}{clang} do not support \href{http://openmp.org/}{OpenMP}). 

We next build a simple minimal example to test that the installation was successful.  
Create a directory called \verb!$HOME/testing!, and inside it create
the file \verb!minimal.cpp!, with the content listed in Listing~\ref{lst1}. A verbatim copy of the above program is also available at \verb!$HOME/qpp/examples/minimal.cpp!.

\begin{lstlisting}[label=lst1, caption={Minimal example}]
// Minimal example
// Source: ./examples/minimal.cpp
#include <iostream>

#include "qpp.h"

int main() {
    using namespace qpp;
    std::cout << "Hello Quantum++!\nThis is the |0> state:\n";
    std::cout << disp(st.z0) << '\n';
}
\end{lstlisting}


Next, compile the file using a C++11 compliant compiler. In the following, we assume that you use \href{https://gcc.gnu.org/}{g++}, but the building instructions are similar for other compilers.
From the directory
\verb!$HOME/testing! type
\begin{verbatim}
g++ -std=c++11 -O3 -Wall -Wextra -pedantic -isystem $HOME/eigen\
 -I $HOME/qpp/include minimal.cpp -o minimal
\end{verbatim}
Your compile command may differ from the above, depending on the C++ compiler and
operating system. If everything went fine, the above command should build an executable \verb!minimal! in \verb!$HOME/testing!, which can be run by typing \verb!./minimal!. The output should be similar to the following:

\begin{lstlisting}[frame=lines, title={Listing~\ref{lst1} output}, keywords={}, numbers=none]
Hello Quantum++!
This is the |0> state:
1 
0 
\end{lstlisting}

In line~5 of Listing~\ref{lst1} we include the main header file of the library \verb!qpp.h! This header file includes all other necessary internal \href{https://github.com/vsoftco/qpp}{Quantum++} header files. In line~10 we display the state $|0\rangle$ represented by the singleton \verb!st.z0! in a nice format using the display manipulator \verb!disp()!.

\section{Data types, constants and global objects\label{sct::data}}
All header files of \href{https://github.com/vsoftco/qpp}{Quantum++} are located inside the \verb!./include! directory. All functions, classes and global objects defined by the library belong to the \verb!namespace qpp!. To avoid additional typing, we will omit the prefix \verb!qpp::! in the rest of this document. We recommend the using of \lstinline!using namespace qpp;! 
in your main \verb!.cpp! file.

\subsection{Data types}
The most important data types are defined in the header file \verb!types.h!. 
We list them in Table~\ref{tbl1}.
\begin{table}[ht]
\begin{adjustwidth}{0in}{0in} 

\begin{tabular}{| l || p{10.85cm} |}
\hline
\verb!idx! & Index (non-negative integer), alias for \verb!std::size_t! \\
\hline
\verb!bigint! & Big integer, alias for \verb!long long int! \\
\hline 
\verb!cplx! & Complex number, alias for \verb!std::complex<double>! \\ 
\hline 
\verb!cmat! & Complex dynamic matrix, alias for \verb!Eigen::MatrixXcd! \\
\hline 
\verb!dmat! & Double dynamic matrix, alias for \verb!Eigen::MatrixXd! \\
\hline 
 \verb!ket! & Complex dynamic column vector, alias for \verb!Eigen::VectorXcd! \\
\hline
\verb!bra! & Complex dynamic row vector, alias for \verb!Eigen::RowVectorXcd! \\
\hline
\verb!dyn_mat<Scalar>! & Dynamic matrix template alias over the field \verb!Scalar!, alias for \verb!Eigen::Matrix<Scalar, Eigen::Dynamic, Eigen::Dynamic>! \\
\hline
\verb!dyn_col_vect<Scalar>! & Dynamic column vector template alias over the field \verb!Scalar!, alias for \verb!Eigen::Matrix<Scalar, Eigen::Dynamic, 1>!\\
\hline
\verb!dyn_row_vect<Scalar>! & Dynamic row vector  template alias over the field \verb!Scalar!, alias for \verb!Eigen::Matrix<Scalar, 1, Eigen::Dynamic>!\\
\hline
\end{tabular}
\caption{User-defined data types}
\label{tbl1}
\end{adjustwidth}

\end{table}

\subsection{Constants}
The important constants are defined in the header file \verb!constants.h! and are listed in Table~\ref{tbl2}.
\begin{table}[ht]
\begin{adjustwidth}{0in}{0in} 

\begin{tabular}{| p{6cm} || l |}
\hline
\verb!constexpr idx maxn = 64;! & Maximum number of allowed qu(d)its (subsystems) \\
\hline 
\verb!constexpr double pi = 3.1415...;! & $\pi$ \\ 
\hline 
\verb!constexpr double ee = 2.7182...;! & $e$, base of natural logarithms \\
\hline
 \verb!constexpr double eps = 1e-12;! & Used in comparing floating point values to zero \\
\hline
\verb!constexpr double chop = 1e-10;!  & Used in display manipulators to set numbers to zero\\
\hline
\verb!constexpr double infty = ...;! & Used to denote infinity in double precision  \\
\hline
\verb!constexpr cplx operator""_i!  \verb!    (unsigned long long int x)!  & User-defined literal for the imaginary number $i:=\sqrt{-1}$\\
\hline
\verb!constexpr cplx operator""_i!  \verb!    (unsigned long double int x)! & User-defined literal for the imaginary number $i:=\sqrt{-1}$ \\
\hline
\verb!cplx omega(idx D)! & $D$-th root of unity $e^{2\pi i/D}$\\
\hline
\end{tabular}
\caption{User-defined constants}
\label{tbl2}
\end{adjustwidth}

\end{table}

\subsection{Singleton classes and their global instances}
Some useful classes are defined as singletons and their instances are globally available, being initialized at runtime in the header file \verb!qpp.h!, before \verb!main()!. They are listed in Table~\ref{tbl3}.
\begin{table}[h!]
\begin{adjustwidth}{0in}{0in} 

\begin{tabular}{| p{8cm} || l  |}
\hline
\verb!const Init& init = Init::get_instance();! & Library initialization \\
\hline
\verb!const Codes& codes = Codes::get_instance();! & Quantum error correcting codes \\
\hline
\verb!const Gates& gt = Gates::get_instance();! & Quantum gates \\
\hline
\verb!const States& st = States::get_instance();! & Quantum states \\
\hline
\verb!RandomDevices& rdevs =! \verb! RandomDevices::get_thread_local_instance();! & Random devices/generators/engines \\
\hline
\end{tabular}
\caption{Global singleton classes and instances}
\label{tbl3}
\end{adjustwidth}

\end{table}

\section{Simple examples\label{sct::examples}}
All of the examples of this section are copied verbatim from the directory \verb!./examples! and are fully compilable. For convenience, the location of the source file is displayed in the first line of each example as a C++ comment. 
The examples are simple and demonstrate the main features of \href{https://github.com/vsoftco/qpp}{Quantum++}. They cover only a small part of library functions, but enough to get the interested user started. For an extensive reference of all library functions, including various overloads, the user should consult the complete reference \verb!./doc/refman.pdf!. See also the rest of the examples (not discussed in this document)  in \verb!./examples/!. for more comprehensive code snippets.

\subsection{Gates and states}
Let us introduce the main objects used by \href{https://github.com/vsoftco/qpp}{Quantum++}: gates, states and basic operations. Consider the code in Listing~\ref{lst2}.

\begin{lstlisting}[label=lst2, caption={Gates and states}]
// Gates and states
// Source: ./examples/gates_states.cpp
#include <iostream>

#include "qpp.h"

int main() {
    using namespace qpp;
    ket psi = st.z0; // |0> state
    cmat U = gt.X;
    ket result = U * psi;

    std::cout << ">> The result of applying the bit-flip gate X on |0> is:\n";
    std::cout << disp(result) << '\n';

    psi = 10_ket; // |10> state
    U = gt.CNOT;  // Controlled-NOT
    result = U * psi;

    std::cout << ">> The result of applying the gate CNOT on |10> is:\n";
    std::cout << disp(result) << '\n';

    U = randU();
    std::cout << ">> Generating a random one-qubit gate U:\n";
    std::cout << disp(U) << '\n';

    result = applyCTRL(psi, U, {0}, {1}); // Controlled-U
    std::cout << ">> The result of applying the CTRL-U gate on |10> is:\n";
    std::cout << disp(result) << '\n';
}
\end{lstlisting}
 
\noindent A possible output is:
 
\begin{lstlisting}[frame=lines, title={Listing~\ref{lst2} output}, keywords={}, numbers=none]
>> The result of applying the bit-flip gate X on |0> is:
0 
1 
>> The result of applying the gate CNOT on |10> is:
0 
0 
0 
1 
>> Generating a random one-qubit gate U:
 -0.251227 - 0.849866i  -0.0204441 - 0.462811i
-0.0716251 + 0.457692i    0.343895 - 0.816777i
>> The result of applying the CTRL-U gate on |10> is:
                    0 
                    0 
 -0.251227 - 0.849866i
-0.0716251 + 0.457692i
\end{lstlisting}

In line~5 of Listing~\ref{lst2} we bring the namespace \verb!qpp! into the global namespace. 

In line~9 we use the \verb!States! singleton \verb!st! to declare \verb!psi! as the zero eigenvector $|0\rangle$ of the $Z$ Pauli operator. In line~10 we use the \verb!Gates! singleton \verb!gt! and assign to \verb!U! the bit flip gate \verb!gt.X!. In line~11 we compute the result of the operation $X|0\rangle$, and display the result $|1\rangle$ in lines~13 and 14.
In line~14 we use the display manipulator \verb!disp()!, which is especially useful when displaying complex matrices, as it displays the entries of the latter in the form $a+bi$, in contrast to the form $(a,b)$ used by the C++ standard library. The manipulator also accepts additional parameters that allows e.g. setting to zero numbers smaller than some given value (useful to chop small values), and it is in addition overloaded for standard containers, iterators and C-style arrays. 

In line~16 we reassign to \verb!psi! the state $|10\rangle$ via the user-defined literal \verb!ket operator"" _ket()!. We could have also used the \href{http://eigen.tuxfamily.org/}{Eigen 3} insertion operator 
 
\begin{lstlisting}[numbers=none]
ket psi(4); // specify the dimension before insertion of elements via <<
psi << 0, 0, 1, 0;
\end{lstlisting}
or the \href{https://github.com/vsoftco/qpp}{Quantum++} library function \verb!mket()!.
In line~17 we declare a gate \verb!U! as the Controlled-NOT with control as the first subsystem, and target 
as the last, using the global singleton \verb!gt!.  In line~18 we declare the ket \verb!result! as the result of applying the Controlled-NOT gate to the state $|10\rangle$, i.e. $|11\rangle$. We then display the result of the computation in lines~20 and 21. 

Next, in line~23 we generate a random unitary gate via the function \verb!randU()!, then in line~27 apply the Controlled-U, with control as the first qubit and target as the second qubit, to the state \verb!psi!. Finally, we display the result in lines~28 and 29.
 
\subsection{Measurements}
Let us now complicate things a bit and introduce measurements. 
Consider the example in Listing~\ref{lst3}.

\begin{lstlisting}[label=lst3, caption={Measurements}]
// Measurements
// Source: ./examples/measurements.cpp
#include <iostream>
#include <tuple>

#include "qpp.h"

int main() {
    using namespace qpp;
    ket psi = 00_ket;
    cmat U = gt.CNOT * kron(gt.H, gt.Id2);
    ket result = U * psi; // we have the Bell state (|00> + |11>) / sqrt(2)

    std::cout << ">> We just produced the Bell state:\n";
    std::cout << disp(result) << '\n';

    // apply a bit flip on the second qubit
    result = apply(result, gt.X, {1}); // we produced (|01> + |10>) / sqrt(2)
    std::cout << ">> We produced the Bell state:\n";
    std::cout << disp(result) << '\n';

    // measure the first qubit in the X basis
    auto measured = measure(result, gt.H, {0});
    std::cout << ">> Measurement result: " << std::get<0>(measured) << '\n';
    std::cout << ">> Probabilities: ";
    std::cout << disp(std::get<1>(measured), ", ") << '\n';
    std::cout << ">> Resulting states:\n";
    for (auto&& it : std::get<2>(measured))
        std::cout << disp(it) << "\n\n";
}
\end{lstlisting}

\noindent A possible output is: 

\begin{lstlisting}[frame=lines, title={Listing~\ref{lst3} output}, keywords={}, numbers=none]
>> We just produced the Bell state:
0.707107 
       0 
       0 
0.707107 
>> We produced the Bell state:
       0 
0.707107 
0.707107 
       0 
>> Measurement result: 1
>> Probabilities: [0.5, 0.5]
>> Resulting states:
0.707107 
0.707107 

-0.707107 
 0.707107 
\end{lstlisting}

In line~11 of Listing~\ref{lst3} we use the function \verb!kron()! to create the tensor product (Kronecker product) of the Hadamard gate on the first qubit and identity on the second qubit, then we left-multiply it by the Controlled-NOT gate. In line~12 we compute the result of the operation $CNOT_{ab}(H\otimes I)|00\rangle$, which is the Bell state $(|00\rangle + |11\rangle)/\sqrt{2}$. We display it in lines~14 and 15. 

In line~18 we use the function \verb!apply()! to apply the gate $X$ on the second qubit of the previously produced Bell state. 
Note that \href{https://github.com/vsoftco/qpp}{Quantum++} uses the C/C++ numbering convention, with indexes starting from zero.
The function \verb!apply()! takes as its third parameter a list of subsystems, and in our case \verb!{1}! denotes the \emph{second} subsystem, not the first. The function \verb!apply()!, as well as many other functions that we will encounter, have a variety of useful overloads, see \verb!doc/refman.pdf! for a detailed library reference. In lines~19 and 20 we display the newly created Bell state.

In line~23 we use the function \verb!measure()! to perform a measurement of the first qubit (subsystem \verb!{0}!) in the $X$ basis. You may be confused by the apparition of \verb!gt.H!, however this overload of the function \verb!measure()! takes as its second parameter the measurement  basis, specified as the columns of a complex matrix. In our case, the eigenvectors of the $X$ operator are just the columns of the Hadamard matrix. As mentioned before, as all other library functions, \verb!measure()! returns by value, hence it does not modify its argument. The return of \verb!measure! is a tuple consisting of the measurement result, the outcome probabilities, and the possible output states. Technically \verb!measure()! returns a tuple of 3 elements
\begin{lstlisting}[numbers=none]
std::tuple<qpp::idx, std::vector<double>, std::vector<qpp::cmat>>
\end{lstlisting}
The first element represents the measurement result, the second the possible output probabilities and the third the output states.
Instead of using this cumbersome type definition, we use the new C++11 \verb!auto! keyword to infer the type of the result \verb!measured! returned by \verb!measure()!.
In lines~24--29 we use the standard \verb!std::get<>()! function to retrieve each element of the tuple, 
then display the measurement result, the probabilities and the resulting output states.

\subsection{Quantum operations}
In Listing~\ref{lst4} we introduce quantum operations: quantum channels, as well as the partial trace and partial transpose operations.

\begin{lstlisting}[label=lst4, caption={Quantum operations}]
// Quantum operations
// Source: ./examples/quantum_operations.cpp
#include <iostream>
#include <vector>

#include "qpp.h"

int main() {
    using namespace qpp;
    cmat rho = st.pb00; // projector onto the Bell state (|00> + |11>) / sqrt(2)
    std::cout << ">> Initial state:\n";
    std::cout << disp(rho) << '\n';

    // partial transpose of first subsystem
    cmat rhoTA = ptranspose(rho, {0});
    std::cout << ">> Eigenvalues of the partial transpose "
              << "of Bell-0 state are:\n";
    std::cout << disp(transpose(hevals(rhoTA))) << '\n';

    std::cout << ">> Measurement channel with 2 Kraus operators:\n";
    std::vector<cmat> Ks{st.pz0, st.pz1}; // 2 Kraus operators
    std::cout << disp(Ks[0]) << "\nand\n" << disp(Ks[1]) << '\n';

    std::cout << ">> Superoperator matrix of the channel:\n";
    std::cout << disp(kraus2super(Ks)) << '\n';

    std::cout << ">> Choi matrix of the channel:\n";
    std::cout << disp(kraus2choi(Ks)) << '\n';

    // apply the channel onto the first subsystem
    cmat rhoOut = apply(rho, Ks, {0});
    std::cout << ">> After applying the measurement channel "
              << "on the first qubit:\n";
    std::cout << disp(rhoOut) << '\n';

    // take the partial trace over the second subsystem
    cmat rhoA = ptrace(rhoOut, {1});
    std::cout << ">> After partially tracing down the second subsystem:\n";
    std::cout << disp(rhoA) << '\n';

    // compute the von-Neumann entropy
    double ent = entropy(rhoA);
    std::cout << ">> Entropy: " << ent << '\n';
}
\end{lstlisting}

\noindent The output of this program is:
 
\begin{lstlisting}[frame=lines, title={Listing~\ref{lst4} output}, keywords={}, numbers=none]
>> Initial state:
0.5   0   0   0.5 
  0   0   0     0 
  0   0   0     0 
0.5   0   0   0.5 
>> Eigenvalues of the partial transpose of Bell-0 state are:
-0.5   0.5   0.5   0.5 
>> Measurement channel with 2 Kraus operators:
1   0 
0   0 
and
0   0 
0   1 
>> Superoperator matrix of the channel:
1   0   0   0 
0   0   0   0 
0   0   0   0 
0   0   0   1 
>> Choi matrix of the channel:
1   0   0   0 
0   0   0   0 
0   0   0   0 
0   0   0   1 
>> After applying the measurement channel on the first qubit:
0.5   0   0     0 
  0   0   0     0 
  0   0   0     0 
  0   0   0   0.5 
>> After partially tracing down the second subsystem:
0.5     0 
  0   0.5 
>> Entropy: 1
\end{lstlisting}

The example should by now be self-explanatory. 
In line~10 of Listing~\ref{lst4} we define the input state \verb!rho! as the projector onto the Bell state $(|00\rangle + |11\rangle)/\sqrt{2}$, then display it in lines~11 and 12.

In lines~15--18 we partially transpose the first qubit, then display the eigenvalues of the resulting matrix \verb!rhoTA!. 
 
In lines~20--22 we define a quantum channel \verb!Ks! consisting of two Kraus operators: $|0\rangle\langle 0|$ and $|1\rangle\langle 1|$, then display the latter. Note that \href{https://github.com/vsoftco/qpp}{Quantum++} uses the \verb!std::vector<cmat>! container to store the Kraus operators and define a quantum channel.

In lines 24--28 we display the superoperator matrix as well as the Choi matrix of the channel \verb!Ks!.

Next, in lines 31--34 we apply the channel \verb!Ks! to the first qubit of the input state \verb!rho!, then display the output state \verb!rhoOut!.

In lines~37--39 we take the partial trace of the output state \verb!rhoOut!, then display the resulting state \verb!rhoA!.

Finally, in lines~42 and 43 we compute the von-Neumann entropy of the resulting state and display it.

\subsection{Timing}
To facilitate simple timing tasks, \href{https://github.com/vsoftco/qpp}{Quantum++} provides a \verb!Timer<>! class that uses internally a \verb!std::steady_clock!. The program in Listing~\ref{lst5} demonstrate its usage.

\begin{lstlisting}[label=lst5, caption={Timing}]
// Timing
// Source: ./examples/timing.cpp
#include <iomanip>
#include <iostream>
#include <vector>

#include "qpp.h"

int main() {
    using namespace qpp;
    std::cout << std::setprecision(8); // increase the default output precision

    // get the first codeword from Shor's [[9,1,3]] code
    ket c0 = codes.codeword(Codes::Type::NINE_QUBIT_SHOR, 0);

    Timer<> t;                           // declare and start a timer
    std::vector<idx> perm = randperm(9); // declare a random permutation
    ket c0perm = syspermute(c0, perm);   // permute the system
    t.toc();                             // stops the timer
    std::cout << ">> Permuting subsystems according to " << disp(perm, ", ");
    std::cout << "\n>> It took " << t << " seconds to permute the subsytems.\n";

    t.tic(); // restart the timer
    std::cout << ">> Inverse permutation: ";
    std::cout << disp(invperm(perm), ", ") << '\n';
    ket c0invperm = syspermute(c0perm, invperm(perm)); // permute again
    std::cout << ">> It took " << t.toc();
    std::cout << " seconds to un-permute the subsystems.\n";

    std::cout << ">> Norm difference: " << norm(c0invperm - c0) << '\n';
}
\end{lstlisting}

\noindent A possible output of this program is:

\begin{lstlisting}[frame=lines, title={Listing~\ref{lst5} output}, keywords={}, numbers=none]
>> Permuting subsystems according to [7, 5, 3, 4, 2, 6, 0, 8, 1]
>> It took 0.000161381 seconds to permute the subsytems.
>> Inverse permutation: [6, 8, 4, 2, 3, 1, 5, 0, 7]
>> It took 0.000104443 seconds to un-permute the subsystems.
>> Norm difference: 0
\end{lstlisting}

In line~11 of Listing~\ref{lst5} we change the default output precision from 4 to 8 decimals after the delimiter. 

In line~14 we use the \verb!Codes! singleton \verb!codes! to retrieve in \verb!c0! the first codeword of the Shor's $[[9,1,3]]$ quantum error correcting code. 

In line~16 we declare an instance \verb!timer! of the class \verb!Timer<>!. In line~17 we declare a random permutation \verb!perm! via the function \verb!randperm()!. In line~18 we permute the codeword according to the permutation \verb!perm! using the function \verb!syspermute()! and store the result . In line~19 we stop the timer. In line~20 we display the permutation, using an overloaded form of the \verb!disp()! manipulator for C++ standard library containers. The latter takes a \verb!std::string! as its second parameter to specify the delimiter between the elements of the container. In line~21 we display the elapsed time using the \verb!ostream operator<<()! operator overload for \verb!Timer<>! instances.

Next, in line~23 we reset the timer, then display the inverse permutation of \verb!perm! in lines~24 and 25. In line~26 we permute the already permuted state \verb!c0perm! according to the inverse permutation of \verb!perm!, and store the result in \verb!c0invperm!. In lines~27 and 28 we display the elapsed time. Note that in line~27 we used directly \verb!t.toc()! in the stream insertion operator, since, for convenience, the member function \verb!Timer<>::toc()! returns a \verb!const Timer<>&!.

Finally, in line~30, we verify that by permuting and permuting again using the inverse permutation we recover the initial codeword, i.e. the norm difference has to be zero. 

\subsection{Input/output}
We now introduce the input/output functions of \href{https://github.com/vsoftco/qpp}{Quantum++}, as well as the input/output interfacing with \href{http://www.mathworks.com/products/matlab/}{MATLAB}. The program in Listing~\ref{lst6} saves a matrix in both \href{https://github.com/vsoftco/qpp}{Quantum++} internal format as well as in \href{http://www.mathworks.com/products/matlab/}{MATLAB} format, then loads it back and tests that the norm difference between the saved/loaded matrix is zero. 

\begin{lstlisting}[label=lst6, caption={Input/output}]
// Input/output
// Source: ./examples/input_output.cpp
#include <iostream>

#include "qpp.h"
#include "MATLAB/matlab.h" // must be explicitly included

int main() {
    using namespace qpp;
    // Quantum++ native input/output
    cmat rho = randrho(256);                 // an 8 qubit density operator
    save(rho, "rho.dat");                    // save it
    cmat loaded_rho = load<cmat>("rho.dat"); // load it back
    // display the difference in norm, should be 0
    std::cout << ">> Norm difference load/save: ";
    std::cout << norm(loaded_rho - rho) << '\n';

    // interfacing with MATLAB
    saveMATLAB(rho, "rho.mat", "rho", "w");
    loaded_rho = loadMATLAB<cmat>("rho.mat", "rho");
    // display the difference in norm, should be 0
    std::cout << ">> Norm difference MATLAB load/save: ";
    std::cout << norm(loaded_rho - rho) << '\n';
}
\end{lstlisting}

\noindent The output of this program is:
 
\begin{lstlisting}[frame=lines, title={Listing~\ref{lst6} output}, keywords={}, numbers=none]
>> Norm difference load/save: 0
>> Norm difference MATLAB load/save: 0
\end{lstlisting}

Note that in order to use the \href{http://www.mathworks.com/products/matlab/}{MATLAB} input/output interface support, you need to explicitly include the header file \verb!MATLAB/matlab.h!, and you also need to have \href{http://www.mathworks.com/products/matlab/}{MATLAB} or \href{http://www.mathworks.com/products/matlab/}{MATLAB} compiler installed, otherwise the program fails to compile. See the Wiki for extensive details about compiling with \href{http://www.mathworks.com/products/matlab/}{MATLAB} support.
 
\subsection{Qudit teleportation}
As mentioned before, \href{https://github.com/vsoftco/qpp}{Quantum++}  treats qubits and qudits on the same footing. Below is a relatively more advanced self-
documented example that implements the teleportation protocol for qudits.

\begin{lstlisting}[label=lst6a, caption={Qudit teleportation}]
// Qudit teleporation
// Source: ./examples/teleport_qudit.cpp
#include <cmath>
#include <iostream>
#include <tuple>
#include <vector>

#include "qpp.h"

int main() {
    using namespace qpp;
    idx D = 3; // size of the system
    std::cout << ">> Qudit teleportation, D = " << D << '\n';

    ket mes_AB = st.mes(D); // maximally entangled state resource

    // circuit that measures in the qudit Bell basis
    cmat Bell_aA =
        adjoint(gt.CTRL(gt.Xd(D), {0}, {1}, 2, D) * kron(gt.Fd(D), gt.Id(D)));

    ket psi_a = randket(D); // random qudit state
    std::cout << ">> Initial state:\n";
    std::cout << disp(psi_a) << '\n';

    ket input_aAB = kron(psi_a, mes_AB); // joint input state aAB
    // output before measurement
    ket output_aAB = apply(input_aAB, Bell_aA, {0, 1}, D);

    // measure on aA
    auto measured_aA = measure(output_aAB, gt.Id(D * D), {0, 1}, D);
    idx m = std::get<0>(measured_aA); // measurement result

    std::vector<idx> midx = n2multiidx(m, {D, D});
    std::cout << ">> Alice measurement result: ";
    std::cout << m << " -> " << disp(midx, " ") << '\n';
    std::cout << ">> Alice measurement probabilities: ";
    std::cout << disp(std::get<1>(measured_aA), ", ") << '\n';

    // conditional result on B before correction
    ket output_m_B = std::get<2>(measured_aA)[m];

    // perform the correction on B
    cmat correction_B =
        powm(gt.Zd(D), midx[0]) * powm(adjoint(gt.Xd(D)), midx[1]);
    std::cout << ">> Bob must apply the correction operator Z^" << midx[0]
              << " X^" << (D - midx[1]) % D << '\n';
    ket psi_B = correction_B * output_m_B;

    // display the output
    std::cout << ">> Bob final state (after correction):\n";
    std::cout << disp(psi_B) << '\n';

    // verification
    std::cout << ">> Norm difference: " << norm(psi_B - psi_a) << '\n';
}
\end{lstlisting}

\noindent The output of this program is:

\begin{lstlisting}[frame=lines, title={Listing~\ref{lst6a} output}, keywords={}, numbers=none]
>> Qudit teleportation, D = 3
>> Initial state:
0.305468 + 0.0132564i
-0.274931 - 0.690466i
-0.537024 - 0.256493i
>> Alice measurement result: 2 -> [0 2]
>> Alice measurement probabilities: [0.111111, 0.111111, 0.111111, 
   0.111111, 0.111111, 0.111111, 0.111111, 0.111111, 0.111111]
>> Bob must apply the correction operator Z^0 X^1
>> Bob final state (after correction):
0.305468 + 0.0132564i
-0.274931 - 0.690466i
-0.537024 - 0.256493i
>> Norm difference: 1.23512e-15
\end{lstlisting}

\subsection{Exceptions}
Most \href{https://github.com/vsoftco/qpp}{Quantum++} functions throw exceptions in the case of unrecoverable errors, such as out-of-range input parameters, input/output errors etc. The exceptions are handled via the class \verb!Exception!, derived from \verb!std::exception!. The exception types are hard-coded inside the strongly-typed enumeration (enum class) \verb!Exception::Type!. If you want to add more exceptions, augment the enumeration \verb!Exception::Type! and also modify accordingly the member  function  \verb!Exception::construct_exception_msg_()!, which constructs the exception message displayed via the overridden virtual function \verb!Exception::what()!. Listing~\ref{lst7} illustrates the basics of exception handling in \href{https://github.com/vsoftco/qpp}{Quantum++}.

\begin{lstlisting}[label=lst7, caption={Exceptions}]
// Exceptions
// Source: ./examples/exceptions.cpp
#include <exception>
#include <iostream>

#include "qpp.h"

int main() {
    using namespace qpp;
    cmat rho = randrho(16); // 4 qubits (subsystems)
    try {
        // the line below throws qpp::exception::SubsysMismatchDims
        double mInfo = qmutualinfo(rho, {0}, {4});
        std::cout << ">> Mutual information between first and last subsystem: ";
        std::cout << mInfo << '\n';
    } catch (const std::exception& e) {
        std::cout << ">> Exception caught: " << e.what() << '\n';
    }
}
\end{lstlisting}

\noindent The output of this program is:

\begin{lstlisting}[frame=lines, title={Listing~\ref{lst7} output}, keywords={}, numbers=none]
>> Exception caught: IN qpp::qmutualinfo(): Subsystems mismatch dimensions!
\end{lstlisting}

In line~10 of Listing~\ref{lst7} we define a random density matrix on four qubits (dimension 16). In line~13, we compute the mutual information between the first and the 5-th subsystem (which does not exist). Line~13 throws an exception of type \verb!qpp::exception::SubsysMismatchDim! exception, as there are only four systems. We next catch the exception in line~16 via the \verb!std::exception! standard exception base class. We could have also used the Quantum++  exception base class \verb!qpp::exception::Exception!, however using the \verb!std::exception! allows the catching of other exceptions, not just of the type \verb!Exception!. Finally, in line~17 we display the corresponding exception message.

\subsection{Classical reversible logic}
\href{https://github.com/vsoftco/qpp}{Quantum++} provides support for classical reversible logic and circuits via 
two classes, \verb!Dynamic_bitset! and \verb!Bit_circuit!. The first is similar to the standard library \verb!std::bitset!, 
with the exception that the length of the bitset can be specified at runtime, whereas the latter is used to describe a classical reversible bit 
circuit and provides the required interface for applying gates, retrieving bit values etc. The example in Listing~\ref{lst8}
is self-explanatory.

\begin{lstlisting}[label=lst8, caption={Classical reversible logic}]
// Reversible classical circuits
// Source: ./examples/reversible.cpp
#include <iostream>

#include "qpp.h"

int main() {
    using namespace qpp;
    std::cout << ">> Classical reversible circuits. ";
    std::cout << "Bits are labeled from right to left,\n   ";
    std::cout << "i.e. bit zero is the least significant bit (rightmost).\n";

    Dynamic_bitset bits{4};                                // 4 classical bits
    std::cout << ">> Initial bitset:\n\t" << bits << '\n'; // display them

    bits.rand(); // randomize the bits
    std::cout << ">> After randomization:\n\t" << bits << '\n'; // display them

    Bit_circuit bit_circuit{bits}; // bit circuit

    std::cout << ">> Apply X_0, followed by CNOT_02, CNOT_13 and TOF_013\n";
    bit_circuit.X(0); // apply a NOT gate on first bit
    bit_circuit.CNOT({0, 2}).CNOT({1, 3}).TOF({0, 1, 3}); // sequence operations

    std::cout << ">> Final bit circuit:\n\t" << bit_circuit << '\n';
    std::cout << ">> 3rd bit: " << bit_circuit.get(2) << '\n';
    std::cout << ">> CNOT count: " << bit_circuit.gate_count.CNOT << '\n';

    bit_circuit.reset(); // resets the circuit
    std::cout << ">> Reseted circuit:\n\t" << bit_circuit << '\n';
    std::cout << ">> CNOT count: " << bit_circuit.gate_count.CNOT << '\n';
}
\end{lstlisting}

\noindent The output of this program is:

\begin{lstlisting}[frame=lines, title={Listing~\ref{lst8} output}, keywords={}, numbers=none]
>> Classical reversible circuits. Bits are labeled from right to left,
   i.e. bit zero is the least significant bit (rightmost).
>> Initial bitset:
	0000
>> After randomization:
	0110
>> Apply X_0, followed by CNOT_02, CNOT_13 and TOF_013
>> Final bit circuit:
	0011
>> 3rd bit: 0
>> CNOT count: 2
>> Reseted circuit:
	0000
>> CNOT count: 0
\end{lstlisting}

\section{Advanced topics\label{sct::advanced}}

\subsection{Aliasing}
Aliasing occurs whenever the same \href{http://eigen.tuxfamily.org/}{Eigen 3} matrix/vector appears on both sides of the assignment operator, and happens because of \href{http://eigen.tuxfamily.org/}{Eigen 3}'s lazy evaluation system. 
Examples that exhibit aliasing: 
\begin{lstlisting}[numbers=none]
mat = 2 * mat;
\end{lstlisting} 
or 
\begin{lstlisting}[numbers=none]
mat = mat.transpose();
\end{lstlisting}
Aliasing \emph{does not} occur in statements like 
\begin{lstlisting}[numbers=none]
mat = f(mat);
\end{lstlisting}
where \verb!f()! returns by value. Aliasing produces in general unexpected results, and should be avoided at all costs.

Whereas the first line produces aliasing, it is not dangerous, since the assignment is done in a one-to-one manner, i.e. each element $(i,j)$ on the left hand side of the assignment operator is solely a function of the the \emph{same} $(i,j)$ element on the right hand side, i.e. $mat(i,j) = f(mat(i,j))$, $\forall i,j$. The problem appears whenever coefficients are being combined and overlap, such as in the second example, where $mat(i,j) = mat(j,i)$, $\forall i,j$. To avoid aliasing, use the member function \verb!eval()! to transform the right hand side object into a temporary, such as 
\begin{lstlisting}[numbers=none]
mat = 2 * mat.eval();
\end{lstlisting}

In general, aliasing can not be detected at compile time, but can be detected at runtime whenever the compile flag \verb!EIGEN_NO_DEBUG! is not set. \href{https://github.com/vsoftco/qpp}{Quantum++} does not  set this flag in debug mode. We highly recommend to first compile your program in debug mode to detect aliasing run-time assertions, as well as other possible issues that may have escaped you, such as assigning a matrix to another matrix of mismatching dimensions etc.

For more details about aliasing, see the official \href{http://eigen.tuxfamily.org/}{Eigen 3} documentation at \url{http://eigen.tuxfamily.org/dox/group__TopicAliasing.html}.

\subsection{Type deduction via \texttt{auto}}
Avoid the usage of \verb!auto! when working with \href{http://eigen.tuxfamily.org/}{Eigen 3} expressions, e.g. avoid writing code like

\begin{lstlisting}[numbers=none]
auto mat = A * B + C;
\end{lstlisting}
but write instead
\begin{lstlisting}[numbers=none]
cmat mat = A * B + C;
\end{lstlisting}
or
\begin{lstlisting}[numbers=none]
auto mat = (A * B + C).eval();
\end{lstlisting}
to force evaluation, as otherwise you may get unexpected results. The ``problem" lies in the \href{http://eigen.tuxfamily.org/}{Eigen 3} lazy evaluation system and reference binding, see e.g. \url{http://stackoverflow.com/q/26705446/3093378} for more details. In short, the reference to the internal data represented by the expression \verb!A * B + C! is dangling at the end of the \verb!auto mat = A * B + C;! statement.

\subsection{Optimizations}
Whenever testing your application, we recommend compiling in debug mode, as \href{http://eigen.tuxfamily.org/}{Eigen 3} run-time assertions can provide extremely helpful feedback on potential issues. Whenever the code is production-ready, you should \emph{always} compile with optimization flags turned on, such as \verb!-O3! (for \href{https://gcc.gnu.org/}{g++}) and \verb!-DEIGEN_NO_DEBUG!. You should also turn on the \href{http://openmp.org/}{OpenMP} (if available) multi-processing flag (\verb!-fopenmp! for \href{https://gcc.gnu.org/}{g++}), as it enables multi-core/multi-processing with shared memory. \href{http://eigen.tuxfamily.org/}{Eigen 3} uses multi-processing when available, e.g. in matrix multiplication. \href{https://github.com/vsoftco/qpp}{Quantum++} also uses multi-processing in computationally-intensive functions.
 
Since most \href{https://github.com/vsoftco/qpp}{Quantum++} functions return by value, in assignments of the form
\begin{lstlisting}[numbers=none]
mat = f(another_mat);
\end{lstlisting}
there is an additional copy assignment operator when assigning the temporary returned by \verb!f()! back to \verb!mat!. As far as we are aware, this extra copy operation is not elided. Unfortunately, \href{http://eigen.tuxfamily.org/}{Eigen 3} does not yet support move semantics, which would have got rid of this additional assignment via the corresponding move assignment operator. If in the future \href{http://eigen.tuxfamily.org/}{Eigen 3} will support move semantics, the additional assignment operator will be ``free", and you won't have to modify any existing code to enable the optimization; the \href{http://eigen.tuxfamily.org/}{Eigen 3} move assignment operator should take care of it for you.

Note that in a line of the form
\begin{lstlisting}[numbers=none]
cmat mat = f(another_mat);
\end{lstlisting}
most compilers perform return value optimization (RVO), i.e. the temporary on the right hand side is constructed directly inside the object \verb|mat|, the copy constructor being elided.

\subsection{Extending Quantum++}
Most \href{https://github.com/vsoftco/qpp}{Quantum++} operate on \href{http://eigen.tuxfamily.org/}{Eigen 3} matrices/vectors, and return either a matrix or a scalar. In principle, you may be tempted to write a new function such as
\begin{lstlisting}[numbers=none]
cmat f(const cmat& A){...}
\end{lstlisting}
The problem with the approach above is that \href{http://eigen.tuxfamily.org/}{Eigen 3} uses \emph{expression templates} as the type of each expression, i.e. different expressions have in general different types, see the official \href{http://eigen.tuxfamily.org/}{Eigen 3} documentation at \url{http://eigen.tuxfamily.org/dox/TopicFunctionTakingEigenTypes.html} for more details. The correct way to write a generic function that is guaranteed to work with any matrix expression is to make the function template and declare the input parameter as \verb!Eigen::MatrixBase<Derived>!, where \verb!Derived! is the template parameter. For example, the \href{https://github.com/vsoftco/qpp}{Quantum++} \verb!transpose()! function is defined as

\begin{lstlisting}
template<typename Derived> 
dyn_mat<typename Derived::Scalar> 
transpose(const Eigen::MatrixBase<Derived>& A){
    const dyn_mat<typename Derived::Scalar>& rA = A.derived();

    // check zero-size
    if (!internal::check_nonzero_size(rA))
        throw Exception("qpp::transpose()", Exception::Type::ZERO_SIZE);

    return rA.transpose();
}
\end{lstlisting}

It takes an \href{http://eigen.tuxfamily.org/}{Eigen 3} matrix expression, line~3, and returns a dynamic matrix over the scalar field of the expression, line~2. In line~4 we implicitly convert the input expression \verb!A! to a dynamic matrix \verb!rA! over the same scalar field as the expression, via binding to a \verb!const! reference, therefore paying no copying cost. We then use \verb!rA! instead of the original expression \verb!A! in the rest of the function. Note that most of the time it is adequate to use the original expression, however there are some cases where you may get a compile time error if the expression is not explicitly cast to a matrix. For consistency, we use this reference binding trick in the code of all \href{https://github.com/vsoftco/qpp}{Quantum++} functions.

\section{Benchmarks\label{sct::bench}}
In this section we compare the performance of \href{https://github.com/vsoftco/qpp}{Quantum++} with two other widely used quantum software 
platforms that allow quantum simulation, namely
IBM's \href{https://qiskit.org}{Qiskit} and the open source \href{http://qutip.org}{QuTiP}. More specifically, we benchmark the time required to perform 
two widely used quantum operations, namely the partial trace and the quantum Fourier transform, respectively, as a function of the number of input 
qubits and number of CPU cores. Note that both vanilla versions of \href{https://qiskit.org}{Qiskit} and \href{http://qutip.org}{QuTiP} do not seem to be 
using parallelization, so for a fair comparison the reader should only compare with the single-threaded curve(s) generated for 
\href{https://github.com/vsoftco/qpp}{Quantum++} . All benchmark plots use logarithmic (base 2) scales for the time scale (expressed in seconds).
We used an 8 core x86-64 Linux machine running Debian 9.5, with an Intel(R) Core(TM) i7-7700K CPU \@ running at 4.20GHz and 16Gb of RAM. 
\href{https://github.com/vsoftco/qpp}{Quantum++} was compiled with \href{https://gcc.gnu.org/}{g++} 6.3, whereas the  
\href{https://qiskit.org}{Qiskit} and \href{http://qutip.org}{QuTiP} simulators were run using \href{https://www.python.org}{Python} 3.5. All  
benchmark code from this section is available online at \url{https://github.com/vsoftco/qpp/tree/master/stress_tests}.
 
\subsection{Partial trace}
\begin{figure}
\centering
      \includegraphics[scale = 0.7]{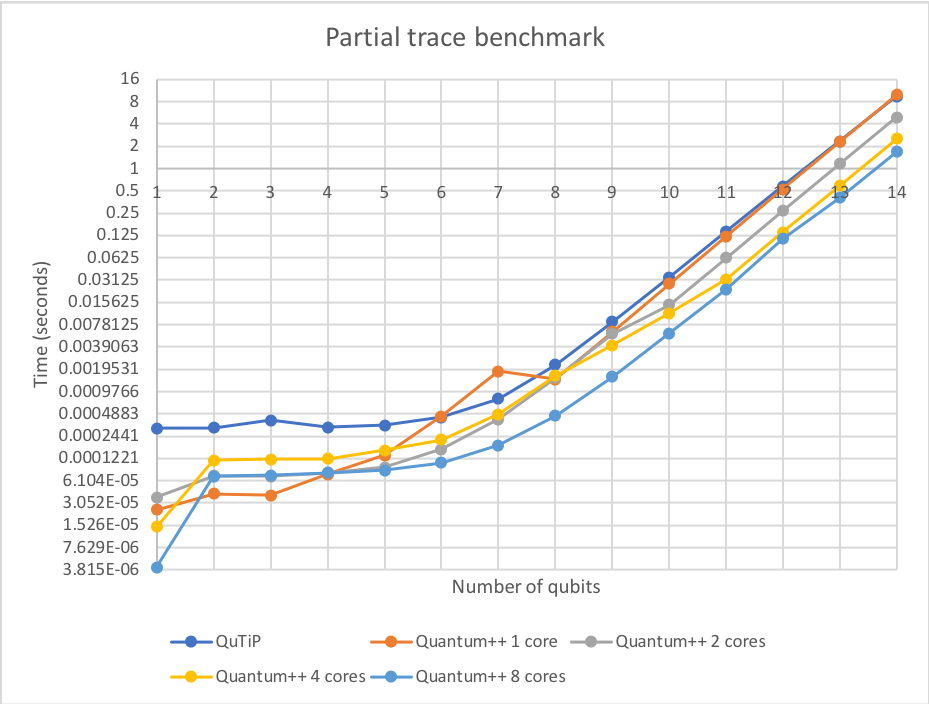}
      \caption{Partial trace on $n$ qubits. Some minor irregularities (spikes) in the plot are most likely due to the fact that the machine we ran the experiments on is not real-time, and the operating system may have performed job scheduling during that time. Note that the single core version of \href{https://github.com/vsoftco/qpp}{Quantum++} over-performs \href{http://qutip.org}{QuTiP}. Remark also that \href{https://github.com/vsoftco/qpp}{Quantum++} scales well with the number of CPU cores.}
      \label{fgr:ptrace}
\end{figure}

In this subsection we benchmark how long it takes to perform a partial trace over the first qubit of an $n$-qubit matrix.
Note that \href{https://qiskit.org}{Qiskit} does not provide a native partial trace function, so we only benchmark against \href{http://qutip.org}{QuTiP}.
The results are displayed in Fig.~\ref{fgr:ptrace}.

\subsection{Quantum Fourier transform}
\begin{figure}
\centering
      \includegraphics[scale = 0.7]{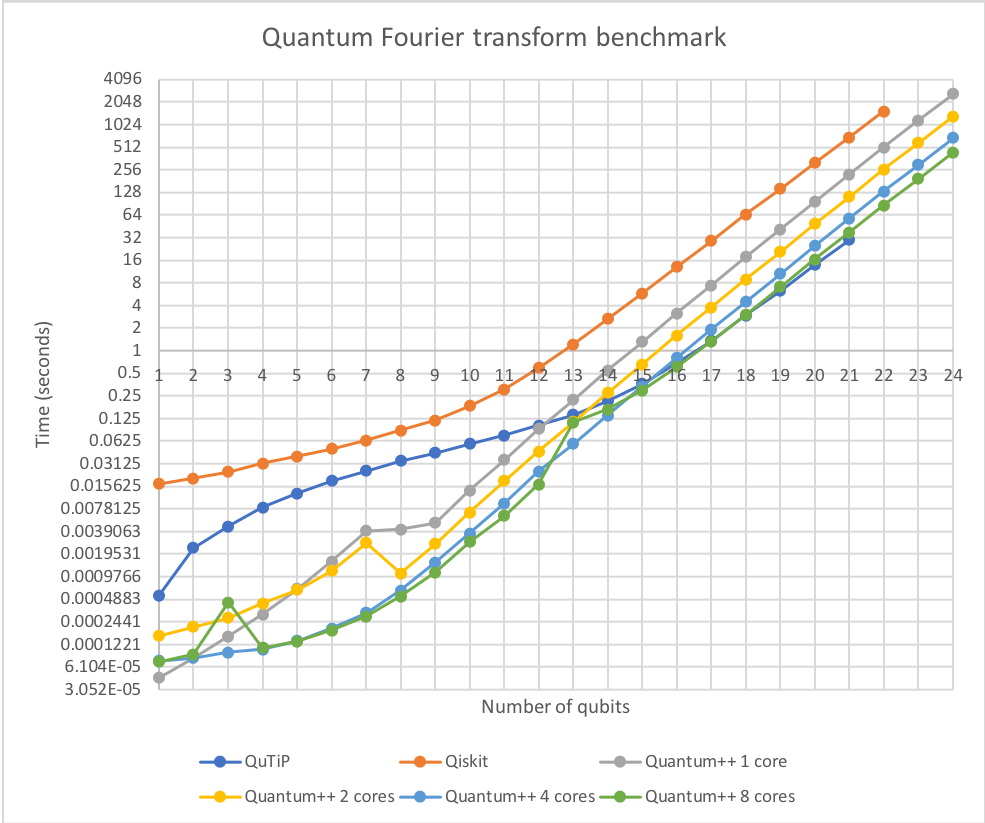}
      \caption{Quantum Fourier transform on $n$ qubits. Some minor irregularities (spikes) in the plot are most likely due to the fact that the machine we ran the experiments is not real-time, and the operating system may have performed job scheduling during that time. Note that the single core version of \href{https://github.com/vsoftco/qpp}{Quantum++} over-performs 
      \href{https://qiskit.org}{Qiskit} by a large margin. \href{http://qutip.org}{QuTiP} seems to be faster in this case than \href{https://github.com/vsoftco/qpp}{Quantum++}. The most likely explanation is that the former uses sparse matrices during computation, whereas the latter does not. However, \href{http://qutip.org}{QuTiP} runs out of memory on our machine after 21 qubits, whereas \href{https://github.com/vsoftco/qpp}{Quantum++} can simulate up to 28 qubits without problems (of course trading the space for longer running time). Remark also that \href{https://github.com/vsoftco/qpp}{Quantum++} scales well with the number of CPU cores.}
      \label{fgr:qft}
\end{figure}

In this subsection we benchmark how long it takes to perform a quantum Fourier transform over the first qubit of an $n$-qubit matrix.
The results are displayed in Fig.~\ref{fgr:qft}.

\subsection{Discussion}
The fundamental data types in \href{https://github.com/vsoftco/qpp}{Quantum++} are non-sparse vectors and matrices. Most computationally-intensive activity performed by the library involves operations on such vectors and matrices. Whenever possible, the task is delegated to the highly-optimized \href{http://eigen.tuxfamily.org/}{Eigen~3} linear algebra library, e.g. when multiplying 2 matrices together. In addition, most loops in the code are parallelized via the \href{http://openmp.org/}{OpenMP} multi-processing library, if the corresponding flag is present at compile time. Those optimizations make \href{https://github.com/vsoftco/qpp}{Quantum++} highly efficient on multiple cores, as the benchmarks in Figures~\ref{fgr:ptrace} and~\ref{fgr:qft} show. 

As one of our referees pointed out, \href{http://qutip.org}{QuTiP} performs relatively poorly for small number of qubits, most likely because of overhead introduced by the Python interpreter. What is surprising is that both \href{http://qutip.org}{QuTiP} and \href{https://qiskit.org}{Qiskit} seem to be single-cored, even though, at least for 
\href{http://qutip.org}{QuTiP}, one would expect aggressive parallelization via the \href{http://www.netlib.org/blas/}{BLAS} library, as mentioned by~\url{http://qutip.org/docs/3.0.1/installation.html\#optimized-blas-librariesl}. Most likely the vanilla version of \href{http://qutip.org}{QuTiP} comes with a \href{http://www.numpy.org}{NumPy} library which is not built against~\href{http://www.netlib.org/blas/}{BLAS}.

\section{Long term maintenance\label{sct::maintain}}
We plan to keep all future releases of \href{https://github.com/vsoftco/qpp}{Quantum++} open source. We will continue to host the project on 
\href{https://github.com}{GitHub} or on an equivalent versioning control system. We will publish new stable releases of the software whenever enough 
improvements or features have been accumulated since the previous stable release. We plan to keep 
\href{https://github.com/vsoftco/qpp}{Quantum++} active and we welcome everyone interested to collaborate.

\section{Conclusions and future directions\label{sct::conclusion}}
As you may have already seen, \href{https://github.com/vsoftco/qpp}{Quantum++} consists mainly of a collection of functions and few classes. There 
is no complicated class hierarchy, and you can regard the \href{https://github.com/vsoftco/qpp}{Quantum++} API as a low/medium-level API. You may 
extend it to incorporate graphical input, e.g. use a graphical library such as \href{http://qt-project.org/}{Qt}, or build a more sophisticated library on top 
of it. We recommend to read the source code and make yourself familiar with the library before deciding to extend it. You should also check the complete 
reference manual \verb!./doc/refman.pdf! for an extensive documentation of all functions and classes.

An interesting future direction is to allow GPU parallelization, however at the time of the writing this was beyond the scope of this project.

\section*{Competing interests}
Dr. Vlad Gheorghiu is the CEO, President and Co-Founder of softwareQ Inc. There are no competing interests with softwareQ Inc.

\section*{Acknowledgements}
I acknowledge financial support from Industry Canada and from the Natural Sciences and Engineering Research Council of Canada (NSERC). 
I thank Sara Zafar Jafarzadeh and Kassem Kalach for carefully reading this manuscript and providing very useful suggestions. 


%
%
%


\begin{thebibliography}{1}

\bibitem{QCsims}
List of {QC} simulators
\newblock \url{http://www.quantiki.org/wiki/List_of_QC_simulators}.

\bibitem{1706.06752}
Roetteler M, Naehrig M, Svore KM, Lauter K. Quantum resource estimates for
  computing elliptic curves discrete logarithms; 	arXiv:1706.06752 [quant-ph], 2017.

\bibitem{NielsenChuang:QuantumComputation}
Nielsen MA, Chuang IL.
\newblock Quantum Computation and Quantum Information.
\newblock 5th ed. Cambridge: Cambridge University Press; 2000.

\end{thebibliography}
\end{document}